\documentclass[11pt]{article} 
\usepackage{mystyle-new}
\usepackage{authblk}
\usepackage[T1]{fontenc}
\usepackage{epsfig,amsmath} 
\usepackage{hepnames,hepunits}
\usepackage{hyperref}
\usepackage{color}
\usepackage{graphicx}
\usepackage{orcidlink}

\definecolor{red}{rgb}{1,0,0}
\def\lesssim{\ \hbox{\raise 2pt \hbox{$<$} \kern -13pt
                     \lower 3pt \hbox{$\sim$}}\ }
\def\greatersim{\ \hbox{\raise 2pt \hbox{$>$} \kern -13pt
                     \lower 3pt \hbox{$\sim$}}\ }

\def\lsim{\mathrel{\rlap{\lower4pt\hbox{\hskip1pt$\sim$}}
    \raise1pt\hbox{$<$}}}                
\def\gsim{\mathrel{\rlap{\lower4pt\hbox{\hskip1pt$\sim$}}
    \raise1pt\hbox{$>$}}}                

\input epsf.tex
\def\desepsf(#1 width #2){\epsfxsize=#2 \epsfbox{#1}}

\newenvironment{tolerant}[1]{\par\tolerance=#1\relax}{ \par }
\usepackage{amsmath,bm}
\usepackage{lineno}

\usepackage{cite,mcite}
\usepackage{tikz}
\usepackage[symbol]{footmisc}

\providecommand{\DOI}[1]{\href{http://dx.doi.org/#1}}

\begin{document}

\title{
Beyond a Year of Sanctions in Science}

\author[1]{
M.~Albrecht~\orcidlink{0009-0006-4513-5322}, 
A.~Ali~\orcidlink{0000-0002-1939-1545},
M.~Barone~\orcidlink{0000-0002-2115-4055},
S.~Brentjes~\orcidlink{0000-0002-8205-8550},
M.~Bona~\orcidlink{0009-0003-7058-7927},
J.~Ellis~\orcidlink{0000-0002-7399-0813},
A.~Glazov~\orcidlink{0000-0002-8553-7338}, 
H.~Jung~\orcidlink{0000-0002-2964-9845},
M.~Mangano~\orcidlink{0000-0002-0886-3789},
G.~Neuneck~\orcidlink{0009-0004-1540-438X}, 
N.~Raicevic~\orcidlink{0000-0002-2386-2290}, 
J.~Scheffran~\orcidlink{0000-0002-7171-3062}, 
M.~Spiro~\orcidlink{0009-0007-1113-1056}, 
P.~van~Mechelen~\orcidlink{0000-0002-8731-9051},
J.~Vigen~\orcidlink{0000-0002-2050-7701}

}

\begin{titlepage} 
\maketitle
\vspace*{-11cm}
\begin{flushright}
\end{flushright}
\vspace*{+10cm}

\begin{abstract}
While sanctions in political and economic areas are now part of the standard repertoire of Western countries (not always endorsed by UN mandates), sanctions  in science and culture in general are new. Historically, fundamental research as conducted at international research centers such as CERN has long been seen as a driver for peace, and the Science4Peace idea has been celebrated for decades. 
However, much changed with the war against Ukraine, and most Western science organizations put scientific cooperation with Russia and Belarus on hold immediately after the start of the war in 2022. In addition, common publications and participation in conferences were banned by some institutions, going against the ideal of free scientific exchange and communication.

These and other points were the topics of an international virtual panel discussion organized by the Science4Peace Forum together with the {\it Natural Scientists Initiative - Responsibility for Peace and Sustainability} (NatWiss e.V.)~\cite{NatWiss} in Germany and the journal {\it Wissenschaft und Frieden} (W\&F)~\cite{WundF} (see the Figure). Fellows from the Hamburg Institute for Peace Research and Security Policy (IFSH)~\cite{IFSH}, scientists collaborating with the large physics research institutes DESY and CERN, as well as from climate and futures researchers were represented on the panel.

In this Dossier we document the panel discussion, and give additional perspectives. \\
~~\\
{\it The authors of the individual sections present their personal reflections, which should not be taken as implying that they are endorsed by the Science4Peace Forum or any other organizations. It is regrettable that some colleagues who expressed support for this document felt that it would be unwise for them to co-sign it.}

\end{abstract}

\end{titlepage}

\begin{figure}[htbp]
   \centering
   \includegraphics[width=0.5\textwidth]{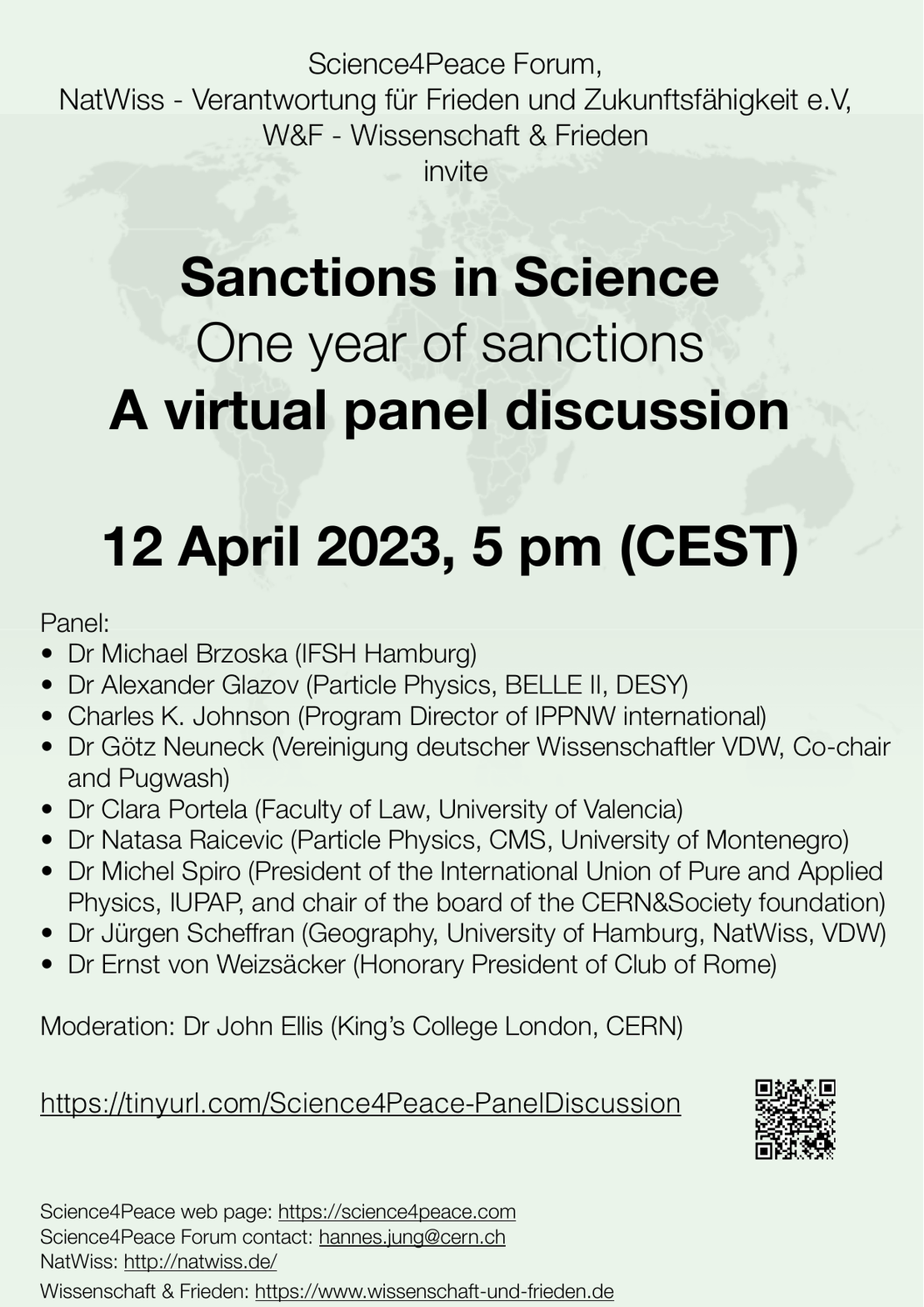} 
   \label{fig:paneldiscussion}
\end{figure}

\section{Introduction} 
\label{Intro}
Following the invasion of the territory of Ukraine by troops of the army of the Russian Federation at the end of February 2022, and the suffering inflicted on many innocent civilians including scientists, the landscape of international scientific collaboration changed greatly. Although many Russian and Belarusian scientists immediately protested against the war \cite{OpenLetterRussianScientistsAgainstWar}, many Western Scientific Institutions launched bans on their historical scientific cooperation with Russian institutions. For example, German Science Organizations \cite{StatementOfAllianceOfGermanScienceOrganizations} recommended freezing all scientific cooperation with Russian State Institutions, and some German research laboratories, such as DESY in Hamburg, went even a step further and banned in addition all common scientific publications and joint participation in scientific conferences \cite{DESY-sanctions1,DESY-sanctions2}.

Scientists at DESY, CERN and elsewhere were very shocked by the war and demanded an immediate stop to this senseless killing of people. While unified in protest against the war, many scientists were also shocked by the immediate reactions of science organizations to put on ice long-standing collaborations, and felt that such actions in the field of science went against the {\it Science for Peace} ideal that had led, in particular, to the foundation of CERN.
After World War II, scientists came together in 1954 and founded CERN, the world's largest research center for particle physics, with the support of their governments and the help of UNESCO. CERN's website states explicitly that one of its missions  is {\it Science for Peace} \cite{CERNmission}. This mission was recognized from the earliest days of CERN. For example, the early history of cooperation between CERN, the Joint Institute for Nuclear Research (JINR) in Dubna and Soviet research institutes from 1955 - 1970 was reviewed in a very interesting article from 1975~\cite{Lock:1975fz}.
More recently, during a Science4Peace seminar in November 2021 \cite{Science4PeaceNov2021}, R. Heuer, a former Director-General of CERN, explained enthusiastically how science at CERN and other laboratories and institutes in different regions of the world contributes as a driver for peace.

Many of us who work at CERN in large experimental collaborations or as visitors enjoyed this spirit of an open scientific community, which allowed  exchanges of scientists and scientific collaboration across borders even during the Cold War. 
We were proud of our international contacts and collaborations, and we were proud to promote scientific collaboration across the world as a driver for peace. 
These principles underpin the solidarity of Western scientists with their Ukrainian colleagues.

As a reaction to the drastic changes in science policy of some science organizations and research laboratories, scientists from DESY and CERN launched on March 3, 2022 an open letter to the DESY directorate to protest against its very strong sanctions imposed on our Russian and Belarusian colleagues~\cite{S4P-letter-to-DESY}, and formed the Science4Peace Forum \cite{S4P-Forum}. After discussions in a wider forum, and being afraid that also in other research laboratories like CERN such strong sanctions could be implemented, a general petition {\it Stop the Escalation Spiral} \cite{SanctionPetition} was launched.  In this petition it is argued that  {\it the sanctions imposed on scientists are counterproductive, they do not put pressure on the Russian government, but make communication among scientists difficult and in some cases impossible. They often affect colleagues who share our condemnation of the war and have endangered their own welfare by expressing their opinions publicly. These sanctions will not help to achieve a ceasefire or resolve the conflict. On the contrary, these measures will isolate Russian and Belarusian scientists and decouple them from international discussions, in science and elsewhere}. The signatories also advocated {\it maintaining scientific cooperation, so as to avoid the useless proliferation of sources of tension that escalate the conflict and extend it to the scientific and personal relations within the physics community}.

After the outbreak of the war, all publications of the big particle physics experiments at CERN were put on hold, and in February 2023 the experiments decided to remove the official affiliations of Russian and Belarusian scientists, replacing them with the phrase {\it affiliated with an institute covered by a cooperation agreement with CERN}\footnote{The original documents of  the decisions of  the experiments are not available publicly, only internally.} \cite{LHCdecisionOnAffilation}. Other big particle physics experiments reacted in different ways, for example the Belle II and BaBar collaborations simply replaced all affiliations by just the ORCID number \cite{BelleII-Statement,BaBar-Statement}.
In September 2022, H. Schopper, the former Director of DESY and also former Director-General of CERN, one of the founding fathers of   the SESAME project in the Middle East,  made very clear statements in an article titled "{\it Science4Peace? More than ever!}"~\cite{CERNcourier-Schopper}.

While scientists are concerned about scientific cooperation and the spirit of international collaboration as a driver for peace when communication is restricted and groups of scientists are excluded from common projects (for further discussion see \cite{SpiegelJung,DUZarticle,W+FBlogHannes}), an even greater worry for everybody is the fear of further escalation and the risk of a nuclear inferno. This led the Science4Peace Forum to launch, together with 14 Nobel Laureates and other well-known scientists, a petition {\it No First Use - Never Any Use of Nuclear Weapons} \cite{NoFirstUsePetition}. 

After more than a year of war against Ukraine, and over a year of sanctions in science, a panel discussion on {\it Sanctions in Science - One Year of Sanctions} \cite{SanctionPanel2023} was organized by the Science4Peace Forum  to recap the consequences of sanctions  and to discuss the future of scientific cooperation. The following Dossier includes contributions to the panel discussion and additional comments on the future of scientific cooperation.

\section{Sanctions in Science - One Year of Sanctions}
In this Section we document the statements and reflections of the panelists at the panel discussion in spring 2023  {\it Sanctions in Science - One Year of Sanctions}. The full video recording is available on \cite{SanctionPanel2023}. 

\subsection{Natasa  Raicevic, CMS, University of Montenegro, Podgorica, Montenegro}
When addressing the problem of sanctions in science, especially in the case of big international scientific collaborations whose results come from the united effort of many scientists from different countries all over the world, the following two points are important to review.
There are examples how sanctions in science were conducted in the past, e.g., their imposition on Yugoslavia in the 1990's,
and an example of cooperation through scientific collaboration between opposing parties in current wars and conflicts, namely the SESAME project.

Thirty years ago, the Federal Republic of Yugoslavia (Serbia and Montenegro) was under embargo and sanctions from UN Security Council because of interference in the civil war in Bosnia. The former Socialist Federal Republic of Yugoslavia had been one of the 12 founding countries of CERN in 1950s. However, it left the Organization in 1961 and at the time when sanctions were imposed it  had the status of Observer to the CERN Council.~\footnote{Serbia has subsequently rejoined CERN as a Member State, and Montenegro has an International Cooperation Agreement (ICA) with CERN.}

As soon as sanctions were imposed, CERN announced that it would not ignore the unanimous will of the international community, and CERN promptly adopted the United Nation's embargo against the Federal Republic of Yugoslavia. 
The details how CERN reacted can be found on the official site: \href{https://home.cern/news/press-release/cern/cern-and-un-embargo-against-serbia-and-montenegro}{\it CERN and UN  embargo against Serbia and Montenegro}.

CERN took all measures to run down the activities of cooperation with Serbia and Montenegro. For example, all data communications using computer networks were shut down. No scientific materials were sent from CERN to Yugoslavia or vice versa. After the embargo was announced, no CERN personnel visited Yugoslavia. CERN also decided not to implement the Agreements of Scientific Cooperation it had signed in 1989 and in 1991, and subsequently discontinued all cooperation with the scientific institutes of Serbia and Montenegro.

Now, thirty years later, very careful and lengthy discussions have taken place within the CERN collaborations in attempt to find the most decent way to apply the sanctions towards our colleagues from Russia and Belarus. After over a year and many rounds of secret voting they arrived at a compromise between several sides with different, strong feelings towards the sanctions.

In the case of the war in Ukraine, CERN did not take such sharp actions toward our colleagues from Russia and Belarus as it had earlier towards those from Yugoslavia (Serbia and Montenegro). CERN softened its actions as long as it could and there were extensive exchanges of opinions in the CERN Council itself and also within and between the collaborations. Appeals from scientists, in particular from the Science4Peace Forum, played an important role in mitigating the sanctions compared to those imposed on Yugoslavia in the 1990s. It should be remembered that the presence, contribution and impact of Russian scientists at CERN was much more significant than the Yugoslav ones, and that sanctions were not being mandated by the UN.

Any sanctions put a country in a difficult situation and, in general, they have the greatest consequences in science and culture. During the recovery, after the political and economic sanctions, the country solves the most urgent problems first, and the collapsed science and culture must wait a long time for their recovery. 

We should highlight all the good examples that were present in science before 2022, and that are still there.
For example, the Synchrotron-light for Experimental Science and Applications in the Middle East (SESAME~\cite{SESAME-home}) regional project should be mentioned. It started working in Jordan officially in 2017, and now represents a true center of science. It was supported by CERN with the aim to use science as a way to learn to work together in the Middle East. Like CERN, it was established with the support of UNESCO as a Science for Peace project. It brings together regional adversaries such as Iran, Israel and Palestine. So, projects are still being developed whose main aims include building scientific and cultural bridges between participating countries, and strengthening mutual understanding and tolerance through international cooperation between people who were recently or are currently in conflict. 

We know this works and there is no reason why should one go and spoil scientific collaborations that are built and function on solid foundations for many years, thanks to the dedication of the scientists who care for it. Such collaborations should not suffer from the many injustices, violations of rights and wars that we continue to witness.

\subsection{Michel Spiro, President of IUPAP and Jens Vigen, Secretary-General of IUPAP}
The International Union of Pure and Applied Physics (IUPAP)~\cite{IUPAP}) has, since its outset hundred years ago, defended the position that no scientists should be barred from participating in conferences or events on the basis of their nationality or their affiliation. This position is clearly reflected in Article 3 in its Articles of Association~\cite{IUPAP-association} (the latest version was adopted by the IUPAP General Assembly on 22 October 2021). 

The text says:
 {\it The purpose of IUPAP is to assist in the worldwide development of physics, to foster {\bf international cooperation} in physics.}

{\it  IUPAP carries out its purpose by {\bf sponsoring international meetings};  .... ; upholding openness, honesty and integrity in the practice, application and promotion of physics; {\bf supporting the free circulation of scientists}; ...}
(our emphasis).

In order to mitigate sanctions against our Russian and Belarusian colleagues, IUPAP at its General Assembly in July 2022 approved the possibility for anyone not actively supporting war and who is committed to democratic principles for resolving disagreements and conflicts to use IUPAP as their affiliation when participating at conferences \cite{IUPAP-affiliation}.

So IUPAP sticks to its principled position, which is that no scientists should be barred from participating in any IUPAP-supported conference or event on the basis of their nationality or affiliation, and that any event where this position is rejected, including rejection of mechanisms that we have formulated to diffuse simmering tensions arising from geo-political conflicts, should not enjoy IUPAP support.

We hope that all colleagues will adhere to the IUPAP position and in that way contribute to foster international cooperation. We will be ready to assist anyone who opts to accept IUPAP affiliation for colleagues affiliated to Russian or Belarusian institutes.

We are ready to extend this mechanism to schools, to publishing, and also to collaborations, with suitable adaptations.

\subsection{G\"otz Neuneck, Co-chair, Federation of German Scientists, German Pugwash Representative and Council Member }
From my perspective, modern science is rooted in humanism and progress for all mankind. Its main principles are objectivity, rational reasoning and international exchange, which must be preserved also in confrontational times. These principles are always endangered. There are  many good historical examples for successful cooperation such as the International Geophysical Year in 1957, the foundation of CERN and DESY,  the International Space Station or the SESAME Project in the Middle East. The scientific community should serve as a bridge across boundaries, as {\it a spearhead of international understanding}, as pointed out by Victor Weisskopf.

On the other hand, scientists should not be naive. Science is not only for international cooperation, but is also a competitive enterprise seeking glory, prestige and national funds. In extreme cases scientific results can be misused for military purposes, exposing the ambivalence of science. The ambivalent nature of scientific knowledge will always exist and can only be mitigated by dialogue, preventive measures, technological assessment and arms control talks. As many political documents show, the global scientific community is more and more challenged by a new geopolitical rivalry between the US, Russia and China. One example is the emerging new arms race between these superpowers. 

The brutal and unlawful war in the Ukraine has not only triggered far-reaching economic sanctions by the European Union against Russia and Belarus, but also led to a freeze of official cooperation with scientific state institutions in Russia and Belarus that support openly  this war of aggression. Russian scientists must exercise the duty of avoiding heated and servile statements to defend an unlawful war against a sovereign country, but instead contribute to a solution. 

Of course, sanctions in science are in general counterproductive in the long run, both for the scientists themselves and for scientific progress as such. Also, state entities should be very careful to assess continuously the individual and scientific implications of these sanctions. Although emphasizing that these sanctions are not applied against single individuals, the collateral damage of official sanctions must be carefully estimated and reversed quickly if the right conditions are met. Western scientists must also talk with their Russian counterparts about the origin, implications and possible resolution of this bloody war in the Ukraine. And there is hope that the relations with Russia can be resurrected once the war is over. 

The Pugwash Conferences on Science and World Affairs have showed for decades that it is not only possible but also absolutely necessary to talk with the other side in a confidential way even on political issues. There are many examples in the history of Pugwash when it was possible to organize a structured dialogue with colleagues in the Eastern bloc or the Middle East on arms control, non-proliferation and disarmament  (restraint, cuts, reductions etc.). Even in times of danger, it still forms a good basis for assessing the consequences of new technologies in security/military affairs using criteria and lessons learnt. Science has a role here, too.

\subsection{Alexander Glazov, Belle II collaboration, DESY, Germany}

High-energy physics collider experiments are vast undertakings involving hundreds or even thousands of physicists from various countries worldwide. For instance, the Belle II experiment is conducted by nearly a thousand physicists from diverse nations, including Japan, several EU countries, China, India, the USA, Ukraine, and Russia.

These collaborations are necessitated by the sheer scale of the projects, both technically and due to the wide array of scientific topics they cover. International collaboration has been crucial since the 1970s and 1980s, a period marked by reduced tensions between Eastern and Western blocs and the disintegration of the Eastern bloc. The collaborative spirit thrived from the 1990s to the 2010s during experiments at prominent institutions like CERN, DESY, Fermilab, KEK, and SLAC.

The primary driving force behind these large experimental collaborations is fundamental research. Collaboration, diversity, openness, and publishing in open-access journals~\cite{arxivScope3} are standard practices. Leadership positions are determined based on these criteria and scientific excellence.

Financial support for high-energy physics research is substantial and primarily sourced from national funding agencies. These agencies benefit significantly from high-profile results, including Nobel awards for discoveries such as the Higgs boson at the LHC and observation of charge-parity violation in $B$-meson decays at SLAC and KEK.

However, the war in Ukraine, initiated by Putin's regime, dealt a blow to this open collaborative environment. The international community's condemnation led to various sanctions imposed by funding agencies, aimed at supporting Ukraine and punishing institutions that did not condemn the war.

Different research collaborations have responded differently to these requests. While vehemently condemning Russian aggression and supporting Ukrainian scientists, collaborations aimed to maintain scientific exchanges and assist Belarusian and Russian researchers who condemned the war. One contentious issue was publication policy. For instance, the Belle II and BaBar collaborations developed a compromise where authors signed papers without affiliating institutes, identified by ORCID number \cite{BelleII-Statement,BaBar-Statement}, meeting both funding agency requirements and the principle of non-discrimination based on nationality.

Sanctions in science have numerous negative effects with little benefit. Most Belarusian and Russian scientists are strongly against the war. Isolating them might push them in the opposite direction, limiting their opportunities for international publication and funding. Additionally:
\begin{itemize}
    \item Belarusian and Russian scientists have made significant contributions to building and running experiments over decades. Discriminating against them is a severe punishment with questionable justification.
    \item There is minimal risk of sharing dual-use technologies; scientists involved in military projects in Russia do not publish in open-access journals.
    \item International collaborations have been crucial for young scientists from Belarus and Russia, fostering mutual benefits for universities with strong research traditions.
    \item Sanctions on Russian institutions mean European collaborators miss out on future projects in Russia, such as NICA~\cite{NICA-collider}, representing lost opportunities for research.
\end{itemize}

One reason for maintaining restrictions on Belarusian and Russian scientists is Ukraine's uncompromising stance. Events allowing Russian participation were boycotted or threatened by Ukraine, understandably given the circumstances. However, it is crucial to look ahead, at the ending of the war and the re-enabling of scientific and humanitarian exchanges. International research in high-energy particle physics has proven to be an ideal platform for this.

\subsection{John Ellis, Theoretical Physicist, King's College London}

My introduction to CERN was as a summer student in 1968. Coming from the cloistered environment of an English university, the international community at CERN was a revelation to me, and I was an instant convert to its mission of ``Science for Peace''~\cite{CERNmission}. At the time, CERN was far less global than it is today but, supported by its Council, had already established relations with scientists in the Soviet Union via the Dubna Joint Institute for Nuclear Research (JINR) and the Institute for High-Energy Physics (IHEP) in Serpukhov~\cite{Lock:1975fz}. Indeed, CERN physicists were working actively on an experiment at IHEP's 70-GeV accelerator, which had the highest energy in the world. This was at the height of the Cold War, but CERN was regarded as a neutral space where scientists from the Soviet Union, Europe and even the US could meet each other and exchange ideas.~\footnote{CERN's role as an East-West meeting-point continued into the 1980s, when discussions during a visit to CERN enabled delegates to the Gorbachev-Reagan summit to make a breakthrough in blocked negotiations.}

It was also in 1968 that the Soviet Union and its satellites invaded Czechoslovakia, but this did not lead to any scientific sanctions. An agreement had been reached earlier in the year to initiate joint CERN-JINR schools of physics, and the invasion delayed the start until 1970, but that was the only significant disruption. These schools went on to introduce many generations of Soviet and West European students to each other, and played an important role in laying a personal basis for East-West collaboration in the years following the collapse of the Soviet Union.~\footnote{It is particularly regrettable that Russian and Belarusian students' access to CERN schools has now been restricted as part of the sanctions discussed below.} Likewise, the CERN experiment at Serpukhov continued, and went on to find the first evidence for rising hadron-hadron cross-sections in 1971.

So, CERN did not impose any scientific sanctions against the Soviet Union following the invasion of Czechoslovakia, nor following the invasion of Afghanistan in 1979. Needless to say, there was also no discussion of sanctions against the US and its allies following their invasions of Afghanistan and Iraq. As discussed in Section 2.1, the only occasion prior to 2022 when CERN implemented any scientific sanctions was in 1992 against the Federal Republic of Yugoslavia (Serbia and Montenegro), aligning itself with resolution 757 of the UN Security Council. For obvious reasons, the UN Security Council did not authorise any sanctions against Russia and Belarus in 2022, so why did CERN implement scientific sanctions? The answer is highly political, and beyond my pay grade.

From 1999 to 2011 I advised successive CERN Directors-General on relations between the organization and many non-Member States, including Russia and Ukraine. In this capacity, ``Science for Peace'' was my personal motto, and it was rewarding to see Indians work alongside Pakistanis, Palestinians work alongside Israelis, and Iranians work alongside Americans, as well as many Middle Eastern, Latin-American and smaller European countries build up their collaborations with CERN. Several of the countries I worked with have become Member States of CERN, including Israel and Serbia, and others have become Associate Members, e.g., India, Pakistan and Ukraine. There were some indications at one stage that Russia might want to become an Associate Member or even a full Member of CERN, but it was not to be.

In parallel with its globalisation, CERN has served as a model for the SESAME project~\cite{SESAME-home} in the Middle East, whose Council brings, in particular, delegates from Iran, Israel and Palestine together around a table to discuss a common scientific project. There are also plans for a similar scientific infrastructure in the Western Balkans called SEEIST~\cite{SEEIST-home}, also inspired by CERN, which would bring together several adversaries in the region including as Albania, Kosovo, Bosnia and Herzegovina, Montenegro and Serbia.

Japan, the US and Russia made key contributions to the construction of the LHC, making possible its scientific successes such as the discovery of the Higgs boson by global teams with over a hundred nationalities. Up until 2022, hundreds of Russian and Belarusian physicists were working alongside their Western colleagues on upgrades of the LHC and its experiments, as well as analysing data. The legal framework for this collaboration is provided by International Co-operation Agreements (ICAs) with Russia, Belarus and JINR that are due to expire in 2024, and the CERN Council has stated its intention not to extend these ICAs.~\footnote{Since JINR is an international scientific organization with several Member States, it should be treated independently from Russian and Belarusian institutes.} This imposition of sanctions against Russian and Belarusian scientists threatens to deprive them of the scientific fruits of their efforts, contrary to the norms on which collaborative research is based. 

Some CERN Member States have gone so far as to forbid their scientists to co-author papers with scientists affiliated with Russian institutions. This embroiled the LHC collaborations in long discussions how to treat Russian and Belarusian authors, as discussed in Section 2.4. For several months there were no LHC papers, followed by a period during which they issued papers with no author lists while tried to find a solution. I advocated simply attaching ORCID identifiers for each author that could provide, when clicked, whatever affiliations and funding information the author wished. This was the solution adopted by the Belle II collaboration at KEK~\cite{BelleII-Statement} (and also for this article), but was not accepted by the LHC collaborations. They decided instead to list authors from Russia and Belarus, but single them out by replacing the names and locations of their institutions by the phrase ``Affiliated with an institute covered by a cooperation agreement with CERN''~\cite{LHCdecisionOnAffilation}. In this way, it is possible at the moment for Russian and Belarusian authors to co-sign CERN papers, but what will happen after the ICAs expire?

The impact on the Russian and Belarusian physicists currently working at CERN will be dire. Many have been based at CERN with their families, some for many years. None bear any personal responsibility for the Russian invasion of Ukraine, most (though not all) oppose it, and many Russian physicists have signed petitions opposing it~\cite{OpenLetterRussianScientistsAgainstWar}. If they are forced to return to Russia, they face potential retribution, perhaps even conscription. What alternatives do they have? Would they consider working on military projects for Russia or some other country such as China, Iran or North Korea? And what of the students and other early-career scientists in Russia who want to do research in high-energy physics? They will lose the opportunity to meet their Western colleagues, and their interests may turn elsewhere, perhaps to military research careers. The point has often been made that scientific sanctions do not deter the Russian regime: in fact, they might even strengthen it, and CERN will lose a generation of potential scientific collaborators.

Scientific sanctions on Russia certainly harm the prospects for future large international science projects. As already mentioned, Russia has been a key partner in the LHC project, but it is difficult to see how this valuable collaboration could be revived for any future accelerator at CERN, such the FCC project. With its determination to ``punish'' Russia for political reasons, the Council of CERN has effectively shot in the foot the organization for which it is responsible. If ``Science for Peace'' is to be more than a slogan, power politics should be kept out of scientific decision-making.

How can we scientists act in the current situation? We should take every opportunity to demonstrate our solidarity with both our Russian and Ukrainian colleagues, and support them in the current situation that is not of their making. We should endeavour to maintain our collaborations with them by informal channels if necessary. We should seek out organisations such as IUPAP~\cite{IUPAP} that may sponsor Russians' participation in scientific events~\cite{IUPAP-affiliation}, or provide affiliations for maintaining scientific collaborations and publishing research results (see Section 2.2). And we should plan ahead for a revival of the ideal of ``Science for Peace'' once the current nightmare comes to an end~\cite{CERNcourier-Schopper}.

\section{Contributions to the discussion from different fields of science}

In this section we document contributions to the discussion on sanctions in science from different fields of science.

\subsection[How can science still cooperate with Russia? Malte Albrecht and J\"urgen Scheffran, Natural Scientists Initiative - Responsibility for Peace and Sustainability]{How can science still cooperate with Russia? Malte Albrecht and J\"urgen Scheffran, Natural Scientists Initiative - Responsibility for Peace and Sustainability~\protect\footnote{Malte Albrecht is a political scientist and chairman of the Natural Scientists Initiative - Responsibility for Peace and Sustainability (NatWiss e.V.~\cite{NatWiss})., J\"urgen Scheffran is a professor of geography at the University of Hamburg and a member of NatWiss.}}

The following text was first published in {\it Frankfurter Rundschau} on July 22, 2022 (and is translated here to English).

"Networks of science with Russia and Ukraine are germ cells of the reconstruction of trust," say Malte Albrecht and J\"urgen Scheffran. In their guest article, they propose concrete steps for a peace-promoting science. 

War cannot be waged with soldiers alone. Scientists all over the world make wars possible with their research. It is not only the natural and technical sciences whose research leads to the development and use of armed drones and automated weapon systems. 

Findings from mass psychology, media and opinion research are central components of military strategy - propaganda is the best-known example. A significant contribution of science to a peaceful world order is to take responsibility for one's own research results. 

Scientific networks with Russia and other states have contributed to a more peaceful world since World War II. They have created spaces for encounter and discussion. Even in times of the greatest polarization and nuclear war threat of the Cold War, there was scientific exchange between East and West. The findings of the Intergovernmental Panel on Climate Change would not be possible without the contributions of Russian research. Without them, we lack understanding the influence of the Arctic and the release of greenhouse gases from Siberian soils on the global climate system and its tipping points.

The destruction of scientific networks does not help. 

The destruction of such networks aims at the political isolation of the Russian government. But the legitimization of the Kremlin is not the task of research networks. 

Therefore, the debate on sanctioning, i.e., suspension of scientific cooperation with Russia, ignores the actual question: How can scientists worldwide contribute to a more peaceful and sustainable world? 

One response to this is the full commitment to civil cooperation, as demanded by the Science4Peace initiative at CERN and DESY in protest against the termination of cooperation with Russia. They can help to contain the instrumentalization of science. Because the debate about the pros and cons of joint research projects ignores the actual problem of international research cooperation: the dependencies of research that counteract the constitutionally guaranteed freedom of science. 

Armaments research is an example of this problem: it is dependent on third-party funds, it must be secret, it leads to scientific block formation, it is part of the war logic. Scientific cooperation with Russia therefore needs the same thing that applies to every other country: a new strategy under the banner of responsibility. 

What can help: 
\begin{itemize}
\item  A practical commitment to peaceful cooperation, for example in the form of civil clauses for projects and allocation of funds. We need every engineer, every social scientist, every economist to tackle the existential challenges of climate change, social inequality and war by peaceful means. 

\item Democratization of academic self-government, especially the highest bodies, to promote decision-making processes from the bottom-up. 

\item Incentives to ensure the peaceful use of research results, in particular of armament-related research. This includes the abolition of patents on innovations that can help solve the most pressing problems of humanity. 

\item A social debate about the financing of science to enable responsible and self-determined research. Research is at the service of the future viability of those who finance it - the citizens. Such networks with Russia and Ukraine are germ cells of the reconstruction of trust and responsibility, of a new security architecture in Europe and the world that involves the interests of all. 
\end{itemize}

Numerous initiatives have spoken out in favor of maintaining the scientific networks with Russia and Ukraine. This cooperation is part of the logic of peace. It is worth being expanded and defended.

\subsection{Sanctions in the sciences as a tool of war-time politics, Sonja Brentjes, Max Planck Institute for the History of Science}

On 22 February 2022, the Russian Federation invaded Ukraine. On 24 February 2022, the foreign minister of the Federal Republic of Germany Annalena Baerbock declared that "we" had woken up to a very different world. Chancellor Scholz seconded her by speaking of a {\it Zeitenwende} (``turning point") and of the first time since World War II that someone tried to change borders in Europe by military means. The German Science Foundation (DFG) hurried to follow suit and stated that "science organizations are restricting or banning scientific cooperation with Russia as part of their scientific tasks and missions." Joybrato Mukherjee, president of the University of Giessen and president of the German Academic Exchange Serviced (DAAD), published a declaration on a blog and then in the journal University World News~\footnote{\href{https://www.universityworldnews.com/post.php?story=20220304085448544}{https://www.universityworldnews.com/post.php?story=20220304085448544}} 
describing how the organization responsible for student and researcher exchange between the Federal Republic of Germany and other states around the globe intended to react to the announced new conditions. Summarizing and simplifying his statement, the DAAD decided to support students, university teachers and researchers from the Ukraine and ban all cooperation with Russia, both without restraint, but with one exception~\footnote{See the second part of this statement.}. This exception concerns the funding of Russian students and scholars within Germany or those who want to study in that country, and opponents of Putin's politics in Russia. When we ask whether this exception was fully implemented during the last year, the policy of sanctions exercised by German science institutions seems to have moved quickly away from the second and third component of the exception. 
German scholars in the humanities, for instance, had to reorient their research to resources held outside the Russian Federation, although this diminished the substance and value of their projects considerably. Science institutions have given instructions even to abstain from publishing together with colleagues from Russian institutions. The quantitative impact of such measures remains unclear, though. A visit to the general DAAD website and its Russia-specific web page indicates no significant changes in the institution's policy, which seems to contradict the president's declaration~\footnote{\href{https://www.daad.ru/de/studieren-forschen-in-russland/}{https://www.daad.ru/de/studieren-forschen-in-russland/}}.
This also seems to be reflected in the published numbers of funded applicants from the Russian Federation and Belarus, which decreased but not as much as implied in the president's statement.~\footnote{\href{https://static.daad.de/media/daad_de/pdfs_nicht_barrierefrei/der-daad/zahlen-fakten/daad-laenderstatistik_115.pdf}{https://static.daad.de/media/daad\_de/pdfs\_nicht\_barrierefrei/der-daad/zahlen-fakten/daad-laenderstatistik\_115.pdf}; \href{https://static.daad.de/media/daad_de/pdfs_nicht_barrierefrei/der-daad/zahlen-fakten/daad-laenderstatistik_113.pdf}{https://static.daad.de/media/daad\_de/pdfs\_nicht\_barrierefrei/der-daad/zahlen-fakten/daad-laenderstatistik\_113.pdf}}
However, the number of funded students and scholars from Ukraine increased significantly. \footnote{\href{https://static.daad.de/media/daad_de/pdfs_nicht_barrierefrei/der-daad/zahlen-fakten/daad-laenderstatistik_114.pdf}{https://static.daad.de/media/daad\_de/pdfs\_nicht\_barrierefrei/der-daad/zahlen-fakten/daad-laenderstatistik\_114.pdf}}

Mukherjee's article contains several statements that indicate a profound shift in the political engagement of leading science organizations of the Federal Republic of Germany since the end of the Cold War. I will limit myself here to a single point: the relationship between science and politics as seen and practiced by the science management. Although there are certain differences between the declarations of the DAAD, the DFG, the MPG or the Humboldt Foundation, all  four institutions with strong engagement in the sciences on the international level currently agree that they should follow the politics of the ruling government in the manner how they as scientific institutions act and implement their mission to promote scientific research, education and cooperation.
Mukherjee formulated this position as follows: {\it These measures mean considerable restrictions in German-Russian scientific 	cooperation and in German-Russian exchange relations. We consider these restrictions to be unavoidable. In science, however, we must be willing to pay this price if we take seriously the fact that, in such a crisis and war situation, our foreign science policy action must be in line with the overall strategy of the German federal government and the European Union.}

In my view such a willingness to subject scientific cooperation to the specific political decisions of a government in office and those of the European Commission is highly questionable. Not only was such an understanding of the relationship between science and politics as well as their respective institutions not subscribed to in the times of the Cold War by all science institutions of the Federal Republic of Germany, as shown in a recent talk by Carola Sachse that focused on the science diplomacy of the MPG with regard to the Soviet Union and China.~\footnote{\href{https://www.youtube.com/watch?v=nk2olVoGC8Y}{https://www.youtube.com/watch?v=nk2olVoGC8Y}} It can also be seen as being in conflict with Article 5 of the German constitution. This article guarantees the freedom of the arts and science, research and teaching. A subjugation of the decisions of science institutions concerning what and whom they sponsor to study, teach or research in cooperation with other scholars and institutions to the specific political interests of those groups that form a government at any given moment in time seems to violate the constitution, a behavior the same article explicitly forbids by obliging these parts of society to uphold the constitution.

Moreover, this willingness to obey the dictum of specific political circles and their views on events is in conflict with previous rhetoric of the very same institutions according to which science needs to serve humanity, peace and the solution of the enormous challenges that we all face. It is beyond doubt that the necessity to support any fight against an illegal war by sanctioning scientific cooperation was not felt by the German science establishment in the major wars since the end of the Cold War, not even those of the twenty-first century. As a contributor to this publication documents, it was rather felt necessary to apply the very same measures of sanctioning the sciences in a combatant country in the moment when Germany was itself party to an illegal war, the war on the Balkans. 

As scholars as well as managers of the science system in the Federal Republic of Germany, and in the face of the role of German sciences between 1933 and 1945, it should rather be our duty to consider carefully how closely, or not, we should be allied to the politics of the day and in which capacity. I think the engagement for peace continues to be our most important duty. Such an engagement demands that we weigh carefully the complexities of each and every situation of political conflict instead of reducing it to the positions of one of the parties involved in it. On the basis of such a balanced, academically sober analysis the great good of scientific cooperation should be defended as, if not a road to peace, then as an avenue for keeping channels of communication open. It should not be sacrificed too easily on the altar of political conformism and gullibility. 

There are further points that need to be considered when such far-reaching changes in the behavior of the science institutions of an entire country take place, for which I have, however, no published information and hence can formulate only questions.

The decision to end the cooperation with Russian and Belarusian academic institutions and to severely curtail the possibilities of carrying out research in those two countries or to publish together with scholars from those two countries was taken by the four leading German science institutions within a surprisingly short period of time, given the usual lengthy process preceding any change in institutional orientation and politics in Germany. How was that possible? Had this decision been prepared for some time, i.e., before the Russian invasion in Ukraine? Did it result from a top-down approach of the German government and the European Union? How was it possible to make those institutions accept this kind of interference in their own institutional principles? If such preemptive decision-making has been undertaken and caused the synchronization of the German science management, the shift in institutional behavior of the four institutions is even more far-reaching than I already assume.

Another major question concerns the reactions of the universities and research institutes to this kind of concerted break up of scientific cooperation with scholars in the Russian Federation and Belarus. Was there any discussion of the possible damage from this political decision? 
Why is it no longer possible to uphold the fundamental principles of scientific research as the guiding lines of science policy? Why are science managers like Mukherjee ready, as he wrote, to pay "a price" instead of reflecting on the consequences of what they mean to do?

One and a half year after this interference into the standard rules of scientific work and international cooperation it is time to discuss what this "price" is and whether there are better ways to fight this or any illegal war.

\section{Experiences under sanctions}
\subsection{Some observations how sanctions against Russia influence scientific life in Russia, Dmitry Kazakov, Dubna}
\begin{itemize}
\item Local conferences: \\
Scientists from countries that joined the sanctions boycott conferences held in Russia.
This also includes online participation, which is not recommended or even forbidden  by some authorities.
This is not true for scientists of other countries like China, India, Egypt, etc.

\item Conferences in the West: \\
Participation of Russian scientists in conferences held in Western countries is hampered.
There are travel difficulties since there are no direct flights to Europe and other
destinations,
and one  to fly through Istanbul or Arab countries. As a result, the journey becomes long and at least twice as expensive.
Another problem is that the bank cards issued by Russian banks do not work abroad any more, so one cannot pay conference fees online and Russian banks cannot transfer money.

Then, the  organisers of some conferences do not allow the use of affiliations with Russian institutes, instead
one has either a blank affiliation or the IUPAP one. This is not accepted by the Russian authorities who support the participation.
Visa applications become more difficult, since many consulates are closed or reduced and one has to apply months in advance in order to obtain a new visa.

\item Joint projects and grants: \\
Joint projects and grants are cancelled or suspended. This affects exchange programs.

\item International Journals: \\
International journals generally continue to accept contributions from Russian institutes  with Russian affiliations. However, their articles are likely to be published behind paywalls (closed access), as sanctioned authors are not in a position to transfer open access fees to publishers based in countries applying the financial sanctions. However, this obstacle to open access does not exist for journals covered by SCOAP3~\cite{scoap3journals}.

\item Participation in experiments: \\
Some collaborations have decided to exclude Russian participants despite their essential financial and intellectual contribution. For instance, the agreement with CERN expires in 2024 and, if it is not prolonged, Russian experimentalists will have to leave CERN experiments. This is inconsistent with the motto {\it Science brings Nations Together}.

\item Experimental hardware: \\
Some experimental equipment produced in Europe cannot be transferred to Russia any more. As a result, the construction of experimental facilities in Russia is partly frozen. For instance, the  launch of the NICA collider at JINR is postponed.
At the same time, part of the equipment produced in Russia is also stored and is not shipped to Europe. This concerns, for instance, magnets for FAIR~\cite{FAIR-collider}.

\item Joint Institute for Nuclear Research (JINR) is an international intergovernmental organisation, like CERN. \\
There are 16 member states and 5 associated members as of today. Nevertheless, JINR is treated  in the same way as Russian organisations, as it is situated in Russia. (CERN is headquartered in Switzerland, but it is not a Swiss organization.). From 2025, the agreement with CERN will expire and JINR experimentalists  will have to leave CERN as well, despite their valuable contributions. This concerns scientists from all JINR member states, not only Russians.
\end{itemize}

\subsection{Negative effects of sanctions, a Russian scientist}
Sanctions in science have numerous negative effects with no benefits at all. They produce frustration and a distance between Russian and Belarusian scientists and their collaborators. 

Does anyone really believe that kicking a few hundred civilians out of DESY and CERN  will stop the confrontation between Russia and Ukraine? The obvious answer is no.

Sanctions on publications and conferences, not allowing young people to apply for summerschools, etc., lead to serious consequences for future communication, with no real justification. CERN claims that it puts measures on institutions but not on individuals. This is hypocrisy. The interruption of international cooperation agreements leads to the loss of association with CERN and the blockage of computing accounts. CERN claims that it does not fight against individuals, but proposes nothing to Russian and Belarusian scientists. It gives just a few positions for some people that are already located at CERN. But these people are less than 10\% of the contributors from Russian and Belarusian institutions. Moreover, not everybody can leave Russia for a long time, due to different family or health reasons. Nothing is proposed for these people.

However, Belarusian and Russian scientists have made significant contributions to building and running experiments over decades. Russia and Belarus contribute in finance, materials and manpower. Data that are collected by the experiments is the common property of international 
collaborations, including Belarusian and Russian scientists. We have the right to analyze these data and to publish results. Is there any real justification for suppressing access to our data? Why can CERN not establish a special status of association without institutions and allow temporarily the use of data and computing facilities?

Two other aspects of the sanctions against Russian and Belarusian scientists:
\begin{itemize}
\item Young people are rejected from the CERN summer school and some other European schools. This means that young people are not allowed to communicate with scientists from the different countries and have no opportunity to learn European culture and to realize that scientists from US and EU are not enemies. Some of them will lead Russian and Belarusian science in 10-15 years. The situation may calm down in some years but they will remember this unjustified rejection. It will jeopardize Russia-US/EU scientific communications for decades.
\item Discrimination in publications is another story. The proposal of Belle II, where all scientists are treated equally using ORCID identifiers, is acceptable. But CERN collaborations completely negate their own diversity and equality statements by putting Russian, Belarusian and JINR scientists in a special cage by writing in their papers: {\it Affiliated with an institute covered by a cooperation agreement with CERN}~\cite{LHCdecisionOnAffilation}. Moreover, even while degrading the role of Russian, Belarusian and JINR scientists in publications, CERN and the collaborations continue to require the same amount of authorship money and service work.
\end{itemize}
Bottom line: there is significant discrimination against Russian and Belarusian scientists on national and geographical grounds. It contradicts to all CERN declarations on equal opportunity, the absence of politics at CERN, etc., produces frustration, and is insulting to Russian and Belarusian scientists. Some of them already refuse to sign papers under these conditions.

\section{Conclusions}

We have reviewed the situation in science after over a year of sanctions imposed on Russians and Belarusians in the science sector. The sanctions imposed  in science have not helped our Ukrainian colleagues, and did not help to end this continuing war, where so many younger and older, brave and clever people with promising future have been killed or seriously injured. 

Following World War II, science was seen as a driver for peace that was the founding principle of CERN, allowing and encouraging communication and collaboration across all borders. This principle was maintained during the Cold War, and was a motivation for generations of young scientists to join this field of research and contribute to the hope for a better world. These principles have been sacrificed on purely political grounds by political leaders and their scientific managers, obviously without having a clear exit strategy, nor clear rules, how, when and under which realistic conditions the sanctions could be lifted. Moreover, it seems that the damage done to the scientific community as a whole, and also to society in general, was not realized. Many contributors to this Dossier mention the frustrations of young scientists who were rejected and not allowed to participate in schools or conferences. And these young scientists will become the leaders of the next scientific generation.

It is important to look to the future and propose ways out of the present dilemma. As mentioned earlier, some of the scientific sanctions contradict the scientific freedom of individual researchers. Indeed, the list of sanctioned institutions was not published by a proper authority but chosen arbitrarily on the basis of their geographical locations. The sanctions are not connected to any actions or support for the war from individuals, and do not apply to researchers in any other part of the world, even if they support Russia's invasion of Ukraine. Advice on the legality of the sanctions is required and, if they are deemed illegal, steps should be taken to cancel the unwarranted restrictions on researchers. 

At the moment, a big issue in physics experiments and collaborations is authorship and how the affiliations of authors are acknowledged. As shown by some experiments, the easiest and most obvious way is to waive all affiliations on scientific publications and give only the ORCID identifiers, where every author can give the information she/he wants to be shown publicly. Such a scenario would at least remove the discrimination present in some of the current author lists.

We should insist that scientific publications and peaceful scientific work should be kept as far away as possible from political discussions and political statements. Science is a universal language that allows people with different backgrounds and different narratives to talk to each other on the basis of equality and respect. The famous conductor Daniel Barenboim said at one of the concerts he gave with his orchestra in Ramallah: "This is not going to bring peace, what it can bring is understanding, patience and courage and curiosity to listen to the narratives of the other"~\cite{Barenboim}. The is perhaps the best description also of the Science4Peace idea.

\vskip 1 cm 
\begin{tolerant}{8000}
\noindent 
{\bf Acknowledgments.} 
We thank all participants in discussions during and after the panel, for their contributions and for their support of this initiative. In particular, we thank Vladimir Lipp for helpful discussions and constructive comments on the text.

\noindent 
\end{tolerant} 
\vskip 0.6cm 

\providecommand{\href}[2]{#2}\begingroup\raggedright\endgroup

\end{document}